\documentclass{article} 
\usepackage[nonatbib, final]{neurips_2023}
\usepackage{subfigure}
\usepackage{graphicx}
\usepackage{tabularx} 
\usepackage{caption}
\usepackage[style=ieee,backend=bibtex]{biblatex}
\usepackage{multirow}
\usepackage{amsmath}

\usepackage[utf8]{inputenc} % allow utf-8 input
\usepackage[T1]{fontenc}    % use 8-bit T1 fonts
\usepackage{hyperref}       % hyperlinks
\usepackage{url}            % simple URL typesetting
\usepackage{booktabs}       % professional-quality tables
\usepackage{amsfonts}       % blackboard math symbols
\usepackage{nicefrac}       % compact symbols for 1/2, etc.
\usepackage{microtype}      % microtypography
\usepackage{xcolor}         % colors

\title{Semi-Supervised Diffusion Model for Brain Age Prediction}

\author{%
 \textbf{Ayodeji Ijishakin}$^{\textnormal{1}}$\thanks{Direct to all correspondence to: ayodeji.ijishakin.21@ucl.ac.uk}, \textbf{Sophie Martin}$^{\textnormal{1}}$, \textbf{Florence Townend}$^{\textnormal{1}}$, \textbf{Federica Agosta}$^{\textnormal{2}}$, \textbf{Edoardo Gioele Spinelli}$^{\textnormal{2}}$, \\ 
 \textbf{Silvia Basaia}$^{\textnormal{2}}$, \textbf{Paride Schito}$^{\textnormal{2}}$, \textbf{Yuri Falzone}$^{\textnormal{2}}$, \textbf{Massimo Filippi}$^{\textnormal{2}}$, \textbf{James Cole}$^{\textnormal{1}}$, \textnormal{and} \hspace{0.2em}\textbf{Andrea Malaspina}$^{\textnormal{1}}$ \\ 
 \\ 
 \hspace*{-4em}$^{\text{1}}$\text{University College London} \\
 \hspace*{-4em}$^{\text{2}}$\text{San Raffelle Scientific Institute}
} 

\begin{document}
\maketitle
\begin{abstract}
Brain age prediction models have succeeded in predicting clinical outcomes in neurodegenerative diseases, but can struggle with tasks involving faster progressing diseases and low quality data. To enhance their performance, we employ a semi-supervised diffusion model, obtaining a 0.83(p<0.01) correlation between chronological and predicted age on low quality T1w MR images. This was competitive with state-of-the-art non-generative methods. Furthermore, the predictions produced by our model were significantly associated with survival length (r=0.24, p<0.05) in Amyotrophic Lateral Sclerosis. Thus, our approach demonstrates the value of diffusion-based architectures for the task of brain age prediction. 
\end{abstract}
\section{Introduction}
Brain age prediction refers to the task of predicting an individual's age using neuroimaging data \cite{Baecker2021MachineApplications}. As individuals age, various brain structures and functions degrade; which lead to negative health outcomes \cite{JamesMolec,AYYBIO, BAMachine,cole2015,deMULTI-MODAL,JamesMolec,slowdown}.
% These include macro-scale brain changes such as: brain volume, and cortical thickness. As well as, microscale degradation to synapses, axons, glial cells, mitochondria and white matter microstructures \cite{deMULTI-MODAL,JamesMolec,slowdown}. 
These changes do not occur at the same rate across people, such that individuals may have the same chronological age but different brain ages \cite{cole2018}.
% This difference is obtained by training a model to learn the association between brain features and age in a healthy cohort, and then applying that model to a new cohort to obtain the difference between their predicted and chronological age, which is often reffered to as the brain predicted age difference (brain-PAD). 
This disparity is particularly present in neurodegenerative diseases such as: Alzheimer's Disease (AD) and Multiple Sclerosis (MS). In such diseases, an older appearing brain has been associated with: disease status, higher risk of mortality, the speed of disease progression, and disease related genetic markers \cite{Ly2020ImprovingDisease,ColePhD2020LongitudinalParadigm,Lowe2016TheDisease,Beheshti2018TheDisease}. 
% Within the brain age paradigm, machine learning algorithms are first trained to predict the ages of healthy controls based on neuroimaging data, in-order to build a model of healthy ageing. These models are often then applied to a disease cohort in-order to measure the brain Predicted Age Difference (brain-PAD), which is calculated as their predicted age minus chronological age. Thus, the brain-PAD serves as a marker of the deviation of individuals from the norm, given their age group. 
%  Other neurodegenerative disease have a much faster progression. An example is: 

However, the brain age paradigm has seen limited application to amyotrophic lateral sclerosis (ALS) \cite{MNDBA}. ALS, is one of the rarest and fastest progressing neurodegenerative diseases; with a median survival time (time from diagnosis to death) of only 3 years \cite{Swinnen2014TheSclerosis}. With such a short time-window for the most significant ageing related brain changes to occur, effective brain age models for ALS need to be particularly sensitive to ageing related neuroanatomical changes. 
% Further, the main localisation of neurodegeneration in ALS is restricted to upper and lower motor neurons, which exacerbates the need for sensitivity \cite{Rothstein2009CurrentSclerosis}. 
One study demonstrated that individuals with cognitive and behavioural symptoms but not motor impairment have an association with a higher brain age; this further outlines the subtlety of the relationship between neuroanatomical ageing and ALS \cite{MNDBA}. Additionally, brain age models are often trained on high quality (research grade) data, which reduces their wider applicability in real-world settings where the quality of data is lower (clinical grade); this is particularly true for rare diseases like ALS.

Generative models may offer a solution, as they have demonstrated great capabilities in capturing latent factors within an image (e.g., ageing) and produce representations which are robust to data quality \cite{Ho,song,DiffAE,Luo,DGM, RepDI,betaVAE,BeatGans}. In particular, diffusion models have demonstrated image synthesis capabilities well beyond previous approaches
% this is exemplified by recent generative art models such as: DALLE-2, Stable Diffusion and mid-journey 
\cite{DALLE2, StableDiffusion, MidJourney}. 
% The data distribution learning capacity of diffusion models lends itself more naturally to image synthesis, than to producing accurate predictive systems. However, recent methodological advances to diffusion models now allow them to condition their image generation on semantically rich representations \cite{DiffAE,StableDiffusion,DALLE2}. 

In this work, we present a novel brain age prediction architecture which leverages the representation learning capabilities of diffusion autoencoders \cite{DiffAE}. Our approach produces representations which are robust to data quality, and conditioned on the neuroanatomical effects of ageing. This results in predictions which are accurate on clinical-grade data and associated with mortality in ALS disease at a level which surpasses non-generative approaches.

\begin{figure*}[]
\centering
    \includegraphics[scale=0.25]{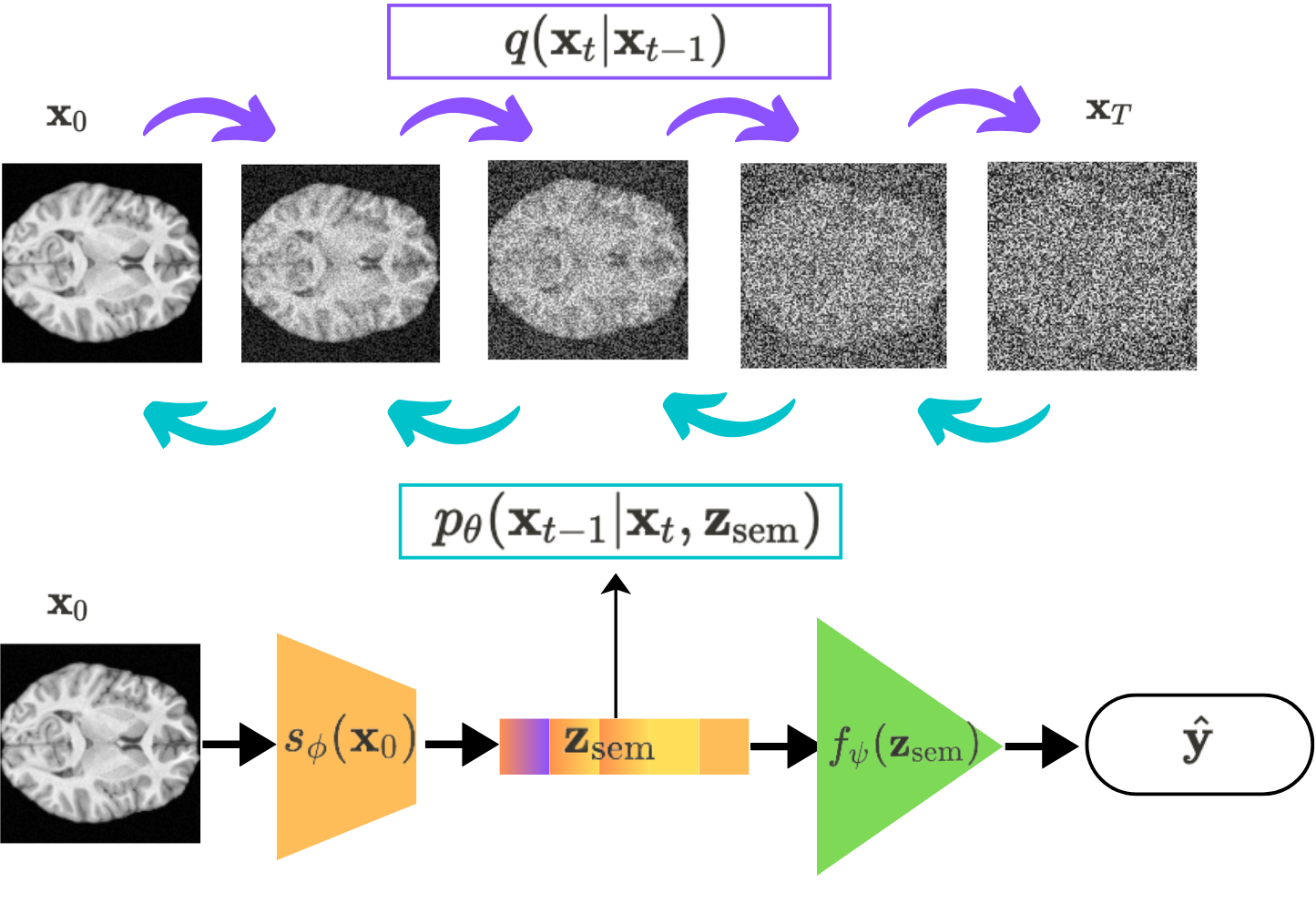}
\caption{The Model architecture. First the image, $\mathbf{x}_{0}$ goes through the forward process $q(\mathbf{x}_{1:T}|\mathbf{x}_{0})$, to produce $\mathbf{x}_{T}$. The reverse process, $p_{\theta}(\mathbf{x_{0:T}}|\mathbf{z}_{\text{sem}})$ then reconstructs the image conditioned on the semantic latent, produced by $s_{\phi}(\mathbf{x}_{0})$. Finally, the age predictor, $f_{\psi}(\mathbf{z}_{\text{sem}})$ predicts an individual's age, $\hat{\mathbf{y}}$. }
\label{Figure1}
\end{figure*}  

\section{Background} 
\subsection{Diffusion Models}
Diffusion models allow us to approximate a data distribution, $q(\mathbf{x}_{0})$, with a learnt distribution, $p_{\theta}({\mathbf{x}}_{0})$, via a latent variable model which takes the form: 
\begin{equation}
    p_{\theta}(\mathbf{x}_{0}) = \int p_{\theta}(\mathbf{x}_{0:T})d\mathbf{x}_{1:T}, \quad \text{where} \quad p_{\theta}(\mathbf{x}_{0:T}) := p_{\theta}(\mathbf{x}_{T})\prod_{t=1}^{T}p_{\theta}^{(t)}(\mathbf{x}_{t-1}| \mathbf{x}_{t})
\end{equation}
Where the latents $\mathbf{x}_{1}, ...,\mathbf{x}_{T}$ share the same dimensions as our training data $\mathbf{x}_{0} \sim q(\mathbf{x}_{0})$ \cite{Ho,song, SOHL}. As denoted above, the transitions between latents are defined as a Markov chain beginning at a prior $p(\mathbf{x}_{T}) = \mathcal{N}(\mathbf{x}_{T}; \textbf{0}, \textbf{I})$ and ending at our model distribution $p_{\theta}(\mathbf{x}_{0})$. We call this joint distribution, $p_{\theta}(\mathbf{x}_{0:T})$, the \textit{reverse process}. In contrast to most latent variable models, the approximate posterior $q(\mathbf{x}_{1:T}|\mathbf{x}_{0})$ is predefined as another Markov chain, that we call the \textit{forward process}. This process destroys the structure of our data, $\mathbf{x}_{0}$, via the gradual addition of Gaussian noise, defined as: 
\begin{equation}
    q(\mathbf{x}_{1:T}|\mathbf{x}_{0}) := \prod^{T}_{t=1} q(\mathbf{x}_{t}|\mathbf{x}_{t-1}), \quad \text{where} \quad q(\mathbf{x}_{t}|\mathbf{x}_{t-1}) := \mathcal{N}(\mathbf{x}_{t};\sqrt{1-\beta_{t}} \mathbf{x}_{t-1}, \beta_{t} \textbf{I})
\end{equation}
Where $\beta_{1}, ..., \beta_{T}$ are scalars which enforce a variance schedule. The reverse process, $p_{\theta}(\mathbf{x}_{0:T})$, denoises our image following the forward process, $q(\mathbf{x}_{1:T}|\mathbf{x}_{0})$, to recover our data distribution. Our model, $p_{\theta}({\mathbf{x}}_{0})$, may be trained by minimising a variational lower bound on negative log likelihood: 
\begin{equation}
    \mathbb{E} \left[ - \log p_{\theta}(\mathbf{x}_{0}) \right] \leq \mathbb{E}_{q} \left[ - \log \frac{p_{\theta}(\mathbf{x}_{0:T})}{q(\mathbf{x}_{1:T}|\mathbf{x}_{0})} \right] = \mathbb{E}_{q} \left[ - \log p_{\theta}(\mathbf{x}_{0:T}) - \log q(\mathbf{x}_{1:T}|\mathbf{x}_{0}) \right]
\end{equation}
At time $t$, the approximate posterior is another Gaussian, $q(\mathbf{x}_{t}|\mathbf{x}_{0}) = \mathcal{N}(\sqrt{\alpha_{t}}\mathbf{x}_{0},(1 - \alpha_{t})\mathbf{I})$, where $\alpha_{t} = \prod_{s=1}^{t}(1 - \beta_{s})$. Thus, the noised image at time $t$ can be expressed as $\mathbf{x}_{t} = \sqrt{\alpha_{t}}\mathbf{x}_{0} + \sqrt{1 - \alpha_{t}}\epsilon$ where $\epsilon \sim \mathcal{N}(\mathbf{0}, \mathbf{I})$. This allows us to simplify our objective, by learning a model, $\epsilon_{\theta}^{(t)}(\mathbf{x}_{t})$, which predicts the sampled noise, $\epsilon$, based on $\mathbf{x}_{t}$ and $t$. This holds provided that our noised image at time $t$ is assumed to be Gaussian with a fixed variance according to $\alpha_{t}$ and a learnable mean admitting the following loss (see \cite{song} for the full derivation): 
\begin{equation}
    \label{Equation-4}
    L := \sum_{t=1}^{T}\mathbb{E}_{\mathbf{x}_{0} \sim q(\mathbf{x}_{0}), \epsilon_{t} \sim \mathcal{N}(\mathbf{0}, \mathbf{I})} \left[ \| \epsilon_{\theta}^{(t)}(\sqrt{\alpha_{t}}\mathbf{x}_{0} + \sqrt{1 - \alpha_{t}}\epsilon_{t}) - \epsilon_{t} \|^{2}_{2}  \right] 
\end{equation} 
\begin{table}[t]
\caption{Results of brain age prediction. The first three rows are different metrics demonstrating the accuracy of the brain age predictions on an external dataset. The final row is the relationship (again Pearson's r) between the brain-PADs and survival in months (time from scan to death) on ALS patients. Significant relationships are displayed in \textbf{bold}.}
\small % Reduce font size
\centering
\begin{sc}
\begin{tabularx}{\linewidth}{Xllllll}
\toprule
& Our Model & DenseNet 121\cite{MIDIBa} & brainageR\cite{Fran} & DeepBrainNet\cite{DeepBrainNet} \\
\midrule
Test R & \textbf{0.83(p<0.001)} & \textbf{0.83(p<0.001)} & \textbf{0.46(p<0.001)} & \textbf{0.81(p<0.001)} \\
Test MAE & 5 years & 8.63 years & 11.7 years & 5.55 years \\
Test R$^{2}$ & 0.65 & 0.12 & -1.14 & 0.56 \\
ALS Survival & \textbf{0.24(p<0.05)} & -0.05 (p=0.61) & 0.17 (p=0.14) & -0.12 (p=0.28) \\
\bottomrule
\end{tabularx} 
\end{sc}
\label{Table1}
\end{table}
\begin{table}[t]
\caption{Results of Two-Sample Kolmogorov-Smirnov tests for significance between the PADs produced by our model and those produced by other models. Significant results are displayed in \textbf{bold}.}
\small % Reduce font size
\centering
\begin{sc}
\begin{tabularx}{\linewidth}{Xllllll}
\toprule
& Our Model & DenseNet 121\cite{MIDIBa} & brainageR\cite{Fran}  & DeepBrainNet\cite{DeepBrainNet} \\
\midrule
Mean & -6.19 & -1.60 & -19.17 & -1.13 \\
Std & 27.29 & 7.74 & 10.97 & 8.06 \\
KS Statistic & & \textbf{0.32(p<0.05)} & \textbf{0.47(p<0.001)} & \textbf{0.31(p<0.05)} \\
\bottomrule
\end{tabularx}
\end{sc}
\label{Table2}
\end{table}
\subsection{Diffusion Autoencoders}
Diffusion autoencoders modify diffusion models by conditioning the reverse process on a latent, $\mathbf{z}_{\text{sem}} \in \mathbb{R}^{d}$ where $\mathbf{x}_{0} \in \mathbb{R}^{D}$ and $d \ll D$. This latent is output by a semantic encoder, $s_{\phi}({\mathbf{x}_{0}}) : \mathbb{R}^{D} \rightarrow \mathbb{R}^{d}$, that is distinct from our noise predictor, $\epsilon_{\theta}^{t}(\mathbf{x}_{t})$. The modified reverse process takes the form: 
\begin{equation}
    p(\mathbf{x}_{0:T}| \mathbf{z_{\text{sem}}}) = p(\mathbf{x}_{T}) \prod_{t=1}^{t=T}p(\mathbf{x}_{t-1}|\mathbf{x}_{t}, \mathbf{z_{\text{sem}}})
\end{equation} 
By conditioning the reverse process on a semantic latent, diffusion autoencoders force the stochasticity contained within the image to be modelled by our forward process, $q(\mathbf{x}_{t}|\mathbf{x}_{0})$ \cite{DiffAE}. Whilst the semantically meaningful information is forced into our latent $\mathbf{z}_{\text{sem}}$, this leads to rich representations which allow for accurate predictions in the present context.

\section{Method} 
Our model fuses a diffusion autoencoder with an age predictor network, $f_{\psi}(\mathbf{z}_{\text{sem}}) : \mathbb{R}^{d} \rightarrow \mathbb{R}^{+}$ which predicts an individual's age based on the latent representation, $\mathbf{z}_{\text{sem}}$ of their brain image, $\mathbf{x}_{0}$. The model takes as input a batch of images sampled from a dataset $q(\mathcal{X}) = \bigl\{\mathbf{x}_{0, i}\bigl\}_{i=1}^{N}$ where $\mathbf{x}_{0, i} \in \mathbb{R}^{D}$. The images are accompanied by age labels, $\mathcal{Y} = \bigl\{\mathbf{y}_{i}\bigl\}$, where $\mathbf{y}_{i} \in \mathbb{R}^{+}$. First the image, $\mathbf{x}_{0}$, goes through the forward process, $q(\mathbf{x}_{1:T}|\mathbf{x}_{0})$, to produce $\mathbf{x}_{T}$. The reverse process, $p_{\theta}(\mathbf{x}_{0:T}|\mathbf{z}_{\text{sem}})$, then reconstructs the image conditioned on the semantic latent, $\mathbf{z}_{\text{sem}}$, produced by $s_{\phi}({\mathbf{x}_{0}})$. Finally, the age predictor, $f_{\psi}(\mathbf{z}_{\text{sem}})$ predicts an individual's age; see Figure \ref{Figure1} for visual intuition.  

Our objective is a sum of two MSE losses. The first loss is modified version of Equation \ref{Equation-4}. With the addition that the noise predictor $\epsilon_{\theta}$ also takes as input $\mathbf{z}_{\text{sem}}$, at every time step. Whilst the second loss is the standard MSE between predicted age and actual age: 
\begin{equation}
    L_{AGE} = \frac{1}{N} \sum^{N}_{i=0} (\mathbf{y}_{i} - \hat{\mathbf{y}_{i}})^{2}
\end{equation}
Thus, the model receives direct supervision from the labels learnt by $f_{\psi}(\mathbf{z}_{\text{sem}})$ and receives self-supervision via its generative arm, learnt by $p_{\theta}(\mathbf{x}_{0:T}|\mathbf{z}_{\text{sem}})$ making for a semi-supervised diffusion model. The model is also semi-supervised in the classic sense of the term as unlabelled are used when training to help condition our latent space. In those instances the image is run through the model in the usual fashion without producing an age prediction $f_{\psi}(\mathbf{z}_{\text{sem}}$). 

\begin{figure*}[t!]
\centering
    \includegraphics[scale=0.5
    ]{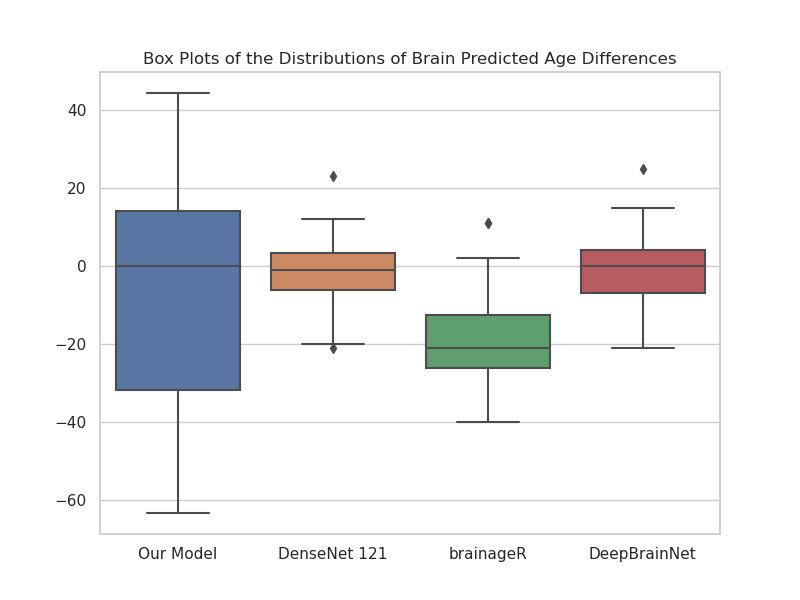}
\caption{A Boxplot of the distribution of brain-PADs produced by different models on our ALS dataset. Our model produces more variation within the brain-PADs of the ALS patients. Two-Sample Kolmogrov Smirnov tests (Table \ref{Table2}) demonstrated that our brain-PADs were significantly different from brain-PADs from the other three approaches .}  

\label{Figure4}
\end{figure*} 

\section{Results}
We trained our model on $4631$ 2D T1w Magnetic Resonance Imaging (MRI) scans from $8$ publicly available datasets (mean age = $56$ years, std = $20.9$ years). For more details concerning the datasets and pre-processing see Appendix \ref{sec:Dataset and Pre-processing}; for details concerning the network design and hardware see Appendix \ref{sec: Net Design and Hardware}. Our test set comprised 473 2D T1w MR images (mean age = 68 years, std = 10 years), 100 \%  of which were clinical grade.

We achieved a test accuracy 0.83(p<0.01) measured as the Pearson's r correlation coefficient between predicted and chronological age (MAE = 5 years, $\text{R}^{2} = 0.65$). Table \ref{Table1} displays our performance compared to state-of-the-art brain age prediction methods, which demonstrates that we are competitive with these approaches. The final row of Table \ref{Table1} shows the correlation between the brain predicted age difference (measured as predicted age minus chronological age) and survival in months in our ALS patient group. This quantity is the biomarker typically used to characterise accelerated ageing \cite{JamesMolec,Baecker2021MachineApplications}. 

The brain predicted age differences (brain-PADs) produced by our model are significantly correlated with survival in months for 72 patients with ALS disease (mean age = 61 years, std = 10.9 years). The other standard approaches did not produce brain-PADs which were significantly associated with this outcome. Table \ref{Table2} shows that the brain-PADs produced by our method were significantly different from the rest of the techniques; see Figure \ref{Figure4} for visual intuition.

% \label{Figure1}
% \end{figure*} 

\section{Discussion\label{sec:discussion}}
In this work, we present a novel diffusion based brain age prediction architecture. To our knowledge, this is the first application of a class of diffusion model to this task. Our results demonstrate that our approach is competitive with state-of-the-art brain age prediction models on a dataset completely composed of clinical-grade data. It exhibited a modest improvement over existing techniques, achieving the joint highest Pearson correlation coefficient (r = 0.83, p < 0.01), the highest coefficient of determination ($\text{R}^{2}$ = 0.65), and the lowest mean absolute error (MAE) of 5 years. 

Most importantly, the ALS brain-PADs produced by our method also capture more variance than non-generative approaches in a statistically significant way; as displayed in the box plots shown in Figure \ref{Figure4}. This variation in brain-PADs was also significantly associated with survival time in ALS, unlike non-generative approaches. This demonstrates that our method is sensitive to the subtle relationship between ALS derived neuroanatomical changes and the ageing process. We posit that our improved performance can be attributed to the capacity of diffusion based architectures, to produce representations which are well conditioned on latent factors within an image, (e.g., the neuroanatomical effects of ageing) and robust to data quality. 

\section{Future Work and Limitations\label{sec:futureandlimis}}
Although the association between the ALS brain-PADs and survival was significant, the results would be better if the effect size were larger. However, the small sample size of ALS patients ($n = 72$) certainly contributed to this. It is also interesting to note that the mean of the brain-PADs was negative ($-6.19$ years), which is the opposite direction of what is expected. Future work should use larger cohorts of ALS patients to see if the effect size of the association between PADs and survival increases; as well as whether the brain-PADs remain mainly negative. The model may also be augmented by including a separate ALS survival objective; but this of course requires more patient data. It may also be extended to process 3D data and should be tested on other diseases such as: AD and MS. We leave this to future work, and treat the present study as a preliminary exposition of our approach to brain age prediction. 

\section{Related Work}
\textbf{Choi et al.} \cite{Choi} used a variational autoencoder to model age conditioned brain topography using positron emission tomography data. The authors found that as they aged individuals with the model, their predicted regional metabolic changes, were associated with their actual follow-up changes. However, this approach was not able to make age predictions as age was required in-order to use the model.

\textbf{Mouches et al.} \cite{Mouches} used an age conditioned autoencoder coupled with a brain age regressor, to produce a model which can produce synthetic brain images of a particular age and perform brain age predictions.  

\textbf{Zhao et al.} \cite{Zhao} similarly combined an age regressor with a variational autoencoder, to model age conditioned latent's  whilst producing brain age predictions. Although both approaches also used MRI data, they were not applied to a patient dataset, so the significance of their approaches for disease groups is unknown. Additionally they were trained solely on research-grade data so are unlikely to generalise to clinical-grade scans. 

\section{Acknowledgements}
This work was supported by funding from the Engineering, and
Physical Sciences Research Council (EPSRC), the UCL Centre for
Doctoral Training in Intelligence, Integrated Imaging in Healthcare
(i4health) and the Motor Neuron Disease (MND) Association.
\newpage 
\printbibliography
\section{Checklist} 
\begin{enumerate}
\item For all authors...
\begin{enumerate}
  \item Do the main claims made in the abstract and introduction accurately reflect the paper's contributions and scope? \textcolor{blue}{[Yes]}
  \item Did you describe the limitations of your work? \textcolor{blue}{[Yes]} see \ref{sec:futureandlimis}. 
  \item Did you discuss any potential negative societal impacts of your work?
    \textcolor{blue}{[Yes]} see \ref{sec:neg_soc}. 
  \item Have you read the ethics review guidelines and ensured that your paper conforms to them? \textcolor{blue}{[Yes]}. 
\end{enumerate}
\item If you are including theoretical results...
\begin{enumerate}
  \item Did you state the full set of assumptions of all theoretical results? \textcolor{gray}{[N/A]}
        \item Did you include complete proofs of all theoretical results? \textcolor{gray}{[N/A]}
\end{enumerate}
\item If you ran experiments...
\begin{enumerate}
  \item Did you include the code, data, and instructions needed to reproduce the main experimental results (either in the supplemental material or as a URL)?\textcolor{red}{[No]} This is because although alot of the data is available from public datasets, individuals must apply to the relevant consortiums in order to have access. We have provided much detail concerning the datasets in Appendix \ref{sec:Dataset and Pre-processing}, such that the relevant applications can be made if someone wishes. We have however included code that can be used to train a model, (see the Abstract). This can then be used to make predictions if the data is acquired. 
  \item Did you specify all the training details (e.g., data splits, hyperparameters, how they were chosen)? \textcolor{blue}{[Yes]} see Appendix \ref{sec: Net Design and Hardware}
        \item Did you report error bars (e.g., with respect to the random seed after running experiments multiple times)?  \textcolor{gray}{[N/A]}
        \item Did you include the total amount of compute and the type of resources used (e.g., type of GPUs, internal cluster, or cloud provider) \textcolor{blue}{[Yes]} see Appendix \ref{sec: Net Design and Hardware}
\end{enumerate}

\item If you are using existing assets (e.g., code, data, models) or curating/releasing new assets...
\begin{enumerate}
  \item If your work uses existing assets, did you cite the creators?  \textcolor{blue}{[Yes]} 
  \item Did you mention the license of the assets?\textcolor{gray}{[N/A]}
  \item Did you include any new assets either in the supplemental material or as a URL? \textcolor{gray}{[N/A]}
  \item Did you discuss whether and how consent was obtained from people whose data you're using/curating? \textcolor{gray}{[N/A]} 
  \item Did you discuss whether the data you are using/curating contains personally identifiable information or offensive content? \textcolor{gray}{[N/A]}
\end{enumerate}

\item If you used crowdsourcing or conducted research with human subjects...
\begin{enumerate}
  \item Did you include the full text of instructions given to participants and screenshots, if applicable? \textcolor{gray}{[N/A]}
  \item Did you describe any potential participant risks, with links to Institutional Review Board (IRB) approvals, if applicable? \textcolor{gray}{[N/A]}
  \item Did you include the estimated hourly wage paid to participants and the total amount spent on participant compensation? \textcolor{gray}{[N/A]}
\end{enumerate}

\end{enumerate}

\newpage 
\section*{Appendix}
\appendix
\section{Potential Negative Societal Impact\label{sec:neg_soc}}
Brain age prediction is primarily used in order to make inferences about cognitive and neuroanatomical decline. Although these predictions may aid in: clinical trial design, disease prognostication and classification. One could imagine brain age prediction being used by bad actors to discriminate against an individual based on their brain age. An example would be an insurance company who would offer higher premiums for individuals with older looking brains. With that being said, we anticipate that the clinical utility of this tool outweighs the likelihood of such events occurring.
 
\section{Dataset and Pre-processing \label{sec:Dataset and Pre-processing}}
We curated one dataset for the task of standard brain age prediction and a separate dataset for ALS survival analyses. Our brain age prediction dataset consisted of 4621 3D structural T1-weighted MR images for training (mean age=56 years, std=20.9 years) and 473 3D structural T1-weighted MR images for testing (mean age=68 years, std=10 years). All individuals included in training were considered to be radiologically normal (of non-disease status). Our data were drawn from 8 publicly
available datasets. These were: the Open Access Series of Imaging
Studies-1, the Australian Imaging,
Biomarker \& Lifestyle Flagship Study of Ageing (AIBL), the Southwest University Adult Lifespan Dataset,
the Alzheimer’s Disease Neuroimaging Initiative (ADNI)
dataset, 
the Dallas Lifespan Brain Study, the Nathan Kline Insti-
tute Rocklands Sample, the National Alzheimer’s Coordinating Center and
the Cambridge Centre for Ageing and Neuroscience study (Cam-CAN). Our ALS dataset comprised 72 3D T1w MR images (mean age = 61 years, std=10.9 years) from the San Rafaele Hospital in Milan. All images underwent the following pre-processing. The ANTS package was used to conduct affine registrations of the images to the MNI 152 brain template, they were then resampled to 130 × 130 × 130 resolution, Simple ITK was used to perform n4 bias field correcttion and HD-BET was used to skull strip the images
\cite{ANTS, SimpleITK, HDBET}.
2D medial axial slices were then taken from the 3D volumes, which had their pixel values normalised to be between 0 and 1, after being resized to 128 × 128 resolution. 

\section{Network Design and Hardware \label{sec: Net Design and Hardware}} 
All model hyperparameters and architectural details were chosen heuristically. Our noise predictor network $\mathbf{\epsilon}^{(t)}_{\theta}$ was a U-Net with the following design. The first convolutional layer expanded the input from a single channel to 32. The tensor stayed at this channel depth for another three convolutional layers. Following this, channel multipliers (2, 3, 4) 
were applied every 3 convolutional layers making
for a channel expansion of (32 → 32 → 32) →  (32 → 32 → 32) → (64 →
64 → 64) → (96 →
96 → 96) → (128 → 128 → 128). Three attention blocks then followed, which received a flattened version of the output; this formed the middle block of the U-Net. The upward path of the U-Net was simply the exact reverse of the downward path with regard to channel expansion. Where the channels were equal across the two U-Net paths, a residual connection sent information across. Group normalisation followed each convolutional layer as well as the SiLU activation
function. Our semantic encoder network, $\mathbf{s}_{\phi}(\mathbf{x}_{0})$ was the  downward path of the U-Net and the middle block. The semantic encoder, output latent representations, $\mathbf{z}_{\text{sem}}$ of dimension $512$. The age predictor network, $\mathbf{f}_{\psi}(\mathbf{z}_{\text{sem}})$, was an MLP which consisted of two linear layers of size: (512 → 128 → 32) following these layers a final layer mapped to the age. Between each layer was a ReLU, batch normalisation and $0.5$ dropout. There were $10.12$ million parameters in total. All training was performed on an Nvidia GeForce RTX 4090 graphics card, with the Adam optimizer using PyTorch lightning \cite{pytl} all model components $\mathbf{\epsilon}^{(t)}_{\theta}(\mathbf{z}_{\text{sem}})$, $s_{\phi}(\mathbf{x}_{0})$ and $f_{\psi}(\mathbf{z}_{\text{sem}})$ were trained simultaneously.

\newpage
\section{Additional Results\label{sec: Additional Results}}

\begin{figure*}[h]
\centering
    \includegraphics[scale=0.3]{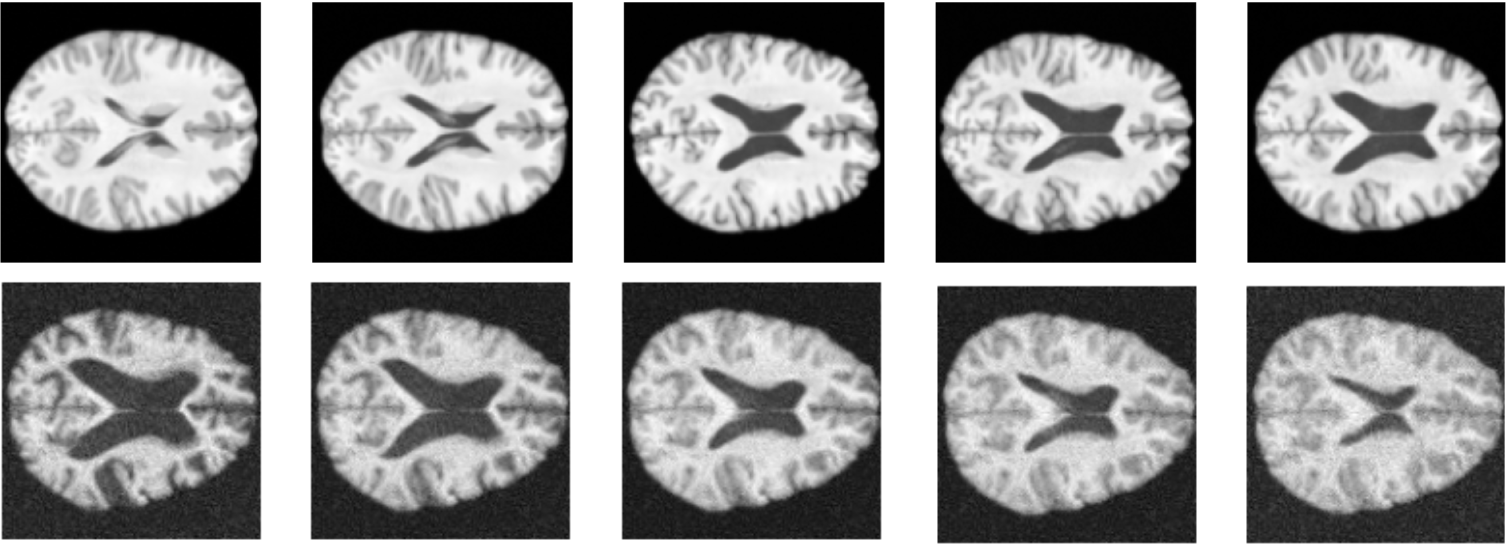}
\caption{Age Interpolations. In the first row, the image in the first column is that of a 25-year-old, we then linearly interpolate between them and a 75-year-old (final column). The second row is the same but reversed, from age 92 to age 20. The difference in data quality, across the two rows, exemplifies the robustness of the model to varied data quality.
}  
\label{Figure3}
\end{figure*} 
\begin{figure*}[h]
\centering
    \includegraphics[scale=0.15]{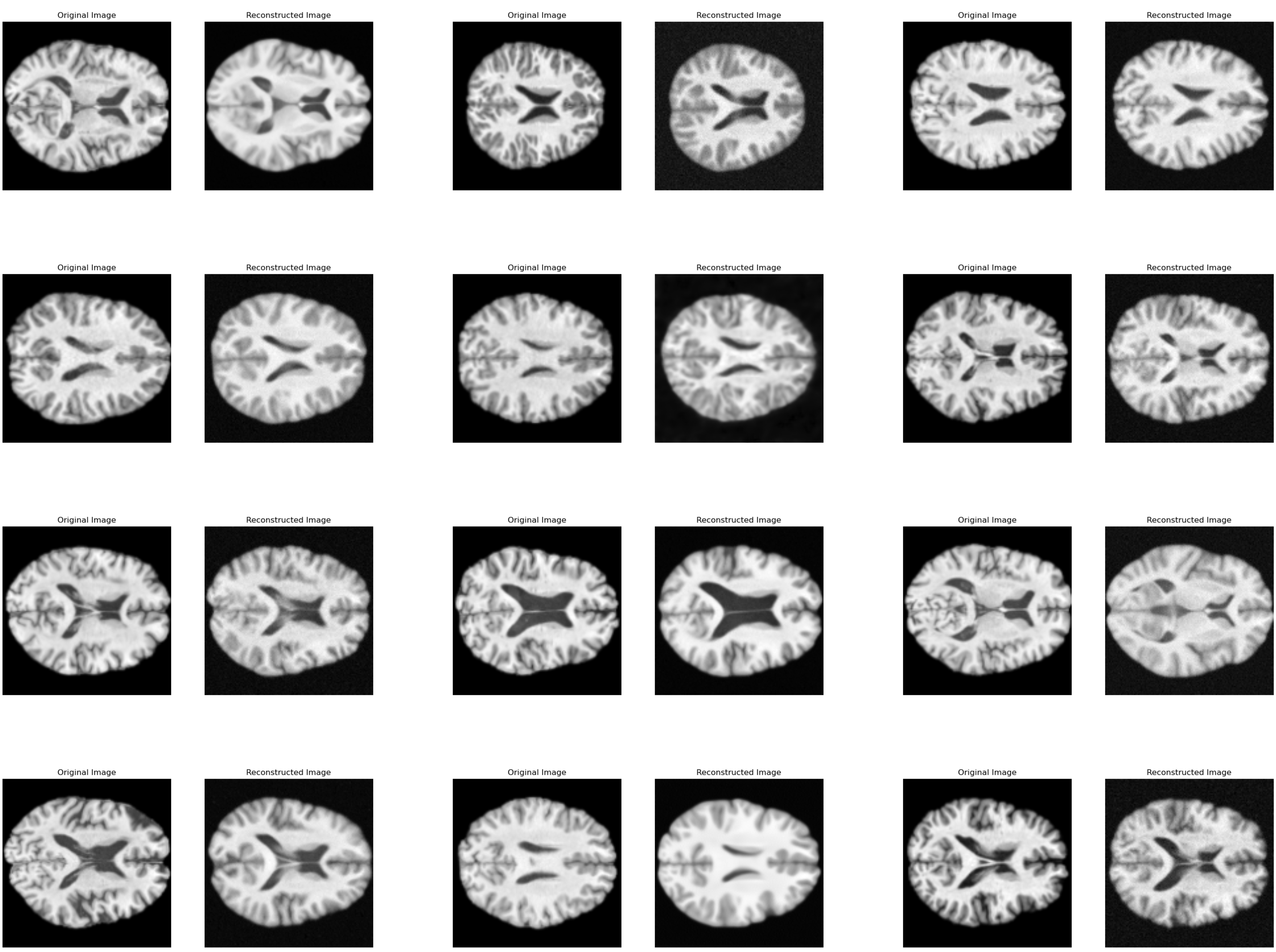}
\caption{Image Reconstructions. The images on the left columns are the original image, and the images on the right columns are their reconstructed version.}   
\label{Figure5}
\end{figure*} 
\end{document}